\documentstyle[aps,epsf,prl,twocolumn]{revtex}

\draft

\begin{document}

\title{Shear-Induced Stress Relaxation in a Two-Dimensional Wet Foam}

\author{John Lauridsen}
\author{Michael Twardos}
\author{Michael Dennin}
\address{Department of Physics and Astronomy}
\address{and Institute for Interfacial and Surface Science}
\address{University of California at Irvine}
\address{Irvine, CA 92697-4575.}

\date{\today}

\maketitle

\begin{abstract}

We report on experimental measurements of the flow behavior
of a wet, two-dimensional foam under conditions of slow,
steady shear. The initial response of the foam is elastic. Above
the yield strain, the foam begins to flow. The flow consists
of irregular intervals of elastic stretch followed by sudden
reductions of the stress, i.e. stress drops. We report on the
distribution of the stress
drops as a function of the applied shear rate. We also
comment on our results in the context
of various two-dimensional models of foams.

\end{abstract}


Foams are ubiquitous in nature \cite{REV} and part of a larger
class of materials which exhibit a type of behavior that has come
to be called ``jamming'' \cite{CWBC98,LN98,TPCSW01}. Loosely
speaking, a ``jammed''
material is one that is unable to flow, typically due to the
packing of the constituent particles. For the case of foams,
which are composed of gas
bubbles separated by fluid walls, the jamming is a consequence
of the topological constraints that develop as the bubbles
press against each other. The existence of a jammed
state contributes to the complex flow behavior of foams.
This flow behavior is one reason that
foams are so interesting from both a fundamental and
applied point of view \cite{REV}. For sufficiently small strains or stresses,
a foam acts as an elastic solid (the jammed state). However, when
the strain (or the stress) exceeds a critical value, known as the
yield strain, a foam begins to flow. Of particular interest is the
flow behavior of foams for small rates of strain. This flow occurs
through ``avalanches'', or sudden nonlinear rearrangements of the
bubbles in the foam that correspond to a decrease in the average
energy, or stress, of the foam. In this paper, we will refer to
these events as ``stress drops''. This type of behavior is common to
many jammed systems where flow occurs in an irregular or
stick-slip type manner when the system is near the transition to
jamming \cite{ITPJamming}. One of the outstanding questions for
flowing foams is the nature of the distribution of the sizes of
the stress drops during this irregular flow.

One reason that this remains an open issue is that predictions for
the distribution of stress drop sizes are somewhat model
dependent. Various models of two-dimensional foams have been used
to study the connection between topological rearrangements and
flow properties as a function of applied rate of strain
\cite{KNN89,KOKN95,OK95,D95,D97,WBHA92,HWB95,JSSAG99}. All of
these simulations predict the same qualitative behavior for the
macroscopic flow for sufficiently small rates of strain that was
described above: elastic behavior below a yield strain and
nonlinear bubble rearrangements above the yield strain. However,
the quantitative predictions for the size distribution and
frequency of these events are model dependent. The dependence is
predominately due to different assumptions
concerning the source of dissipation in foams and treating foams
with different degrees of ``dryness''. Distinguishing between these
different models, and their underlying assumptions, experimentally
is an important step in our understanding of foam rheology.

Two-dimensional foams may be characterized by the area fraction of
gas, $\phi$. For $\phi < 0.84$, foams ``melt'' into a froth of
exclusively circular bubbles. Near this transition, where bubbles
are predominately circular, a foam is said to be ``wet''. In the
limit $\phi$ approaches 1, the bubbles become polygonal with
infinitely thin walls, and a foam is said to be ``dry''. The simulations
studied by Kawasaki, et al. \cite{KNN89,KOKN95,OK95} are
based on the vertex models of foams. This model assumes infinitely
thin walls and only treats the dynamics of the vertices. As such,
it is applicable to dry foams and unlikely to apply
to the wet-foam system studied in this work.
The simulations of Weaire, et al. \cite{WBHA92,HWB95}
focus on the behavior of foams under quasi-static, extensional
flow. These are applicable to wet foams and, because they
are quasistatic, contain no dissipation. These simulations
measure the number of T1 events. A T1 event is a topological
change that involves the switching of neighbors between bubbles.
Reference~\cite{HWB95} reports events that include
large numbers of T1 events, implying a distribution of
stress drop with a power-law like behavior. In contrast, the
bubble model of Durian, which is also applicable to a wet foam,
predicts power-law distributions with an exponential cutoff
\cite{D95,D97}. This model includes viscous dissipation between
bubbles. Further work on this model demonstrated that the
exponential cutoff increases as one approaches the melting point
\cite{TSDKLL99}. However, the dependence on area fraction is
weak, and true power-law behavior was never observed, even extremely
close to the melting point \cite{TSDKLL99}.
Finally, simulations of a q-potts model of a sheared foam
\cite{JSSAG99} did not assume a particular form for the
dissipation. This work suggests that the distribution of topological
rearrangements is not power-law like; however, the distribution
of energy drops may be consistent with power-law behavior \cite{JSSAG99}.

In order to distinguish between these various theoretical models,
there has been limited experimental work on two-dimensional model
foams \cite{DK97,KE99} and three-dimensional foams \cite{GD95}. The
work in Ref. \cite{GD95} made indirect measurements of the bubble
rearrangements and concluded that only local rearrangements
occurred during flow. The work in Ref. \cite{DK97} used Langmuir
monolayers to form two-dimensional foam. This work directly
measured the distribution of T1 events and concluded that the size
distribution was not consistent with a power-law. In contrast, the
work in Ref. \cite{KE99} used a single layer of bubbles trapped
between a glass plate and a water surface. This work focused on
the bubbles whose topological class had changed. These
measurements suggest that very large events can occur.

The apparent discrepancies in these measurements may not be as
severe as it first appears. Detailed simulations of various models
show that the size of T1 events and the number of bubbles involved
in a rearrangement are not necessarily correlated
\cite{JSSAG99,TSDKLL99}. Also, it is possible that the number
of T1 events is not correlated with the size of a stress drop
\cite{JSSAG99}. This is reasonable when one considers that 
not all T1 events will relieve an equivalent amount of stress.
Therefore, to
address the question of the scaling of the size distribution, it
is critical to have direct measurements of the stress or energy
drops. We have accomplished this for the case of another model,
two-dimensional foam: a bubble raft \cite{AK79,MGC89}.

A bubble raft consists of a single layer of bubbles placed on the
surface of water. Bubble rafts have been used to model the
flow behavior of amorphous solids \cite{AK79,MGC89}. The bubble rafts are
an ideal system for the study of two-dimensional foams for a
number of reasons. They allow for both direct measurements of the
macroscopic properties of the foam and the bubble dynamics. There
are no confining glass plates, which can add complications as the
shear rate is increased. Finally, one can control both the degree
of order in the foam (by varying the distribution of bubble sizes)
and the density of the foam with great precision. In this paper,
we will report on the flow behavior of a disordered bubble raft
with $\phi \approx 0.9$.

We generate flow of the bubble raft using a two-dimensional
Couette viscometer that is described in detail in Ref. \cite{app}.
The apparatus consists of two concentric cylinders oriented
vertically. Water is placed between the cylinders, and the upper
surface of the water is free. The outer cylinder consists of 12
individual pieces, so it can be expanded and compressed to adjust
the density of the bubble raft. The working radius of the outer
cylinder was 7.43 cm. The inner cylinder is two pieces:
a solid cylinder placed in the water and a second piece that is in
contact with the bubble raft and fits over the solid cylinder. The
second piece is hung by a torsion wire and is used to measure the
stress, $\sigma$, on the inner rotor due to the bubble raft.
It is important to note that this is
a ``two-dimensional stress'' given by the force on the inner rotor
in the tangential direction divided by the circumference of the
rotor. Therefore, in terms of the torque, $\tau$ on the rotor and
the radius of the rotor $r$, the stress is given by
$\sigma = \tau/(2\pi r^2)$. The radius of the inner cylinder was
3.84 cm. A constant rate of strain is applied to
the system by rotating the outer cylinder at a constant angular
speed in the range 0.0005 rad/s to 0.01 rad/s

The stress on the inner cylinder is determined from the angular
displacement of the torsion wire supporting the inner cylinder.
The angular displacement was measured using magnetic flux. A coil
was attached to the torsion wire and suspended within a
high-frequency magnetic field. The induced voltage was used to
determine the angle of the coil. Typical values of the angle during
shear ranged from $22^{\circ}$ to $40^{\circ}$, corresponding to
stresses in the range of 4 - 23 dyne/cm. The resolution in stress was
set by our resolution in angular measurement and the
torsion constant, $\kappa = 570\ {\rm dyne\ cm/rad}$. The
voltage signal was digitized using a 12-bit A to D converter in the
computer. As a test of the noise level in the signal, the stress
was monitored as a function of time without shear. In the absence
of shear, the noise in
the measured stress signal was at the level of the lowest bit
in digitized signal, corresponding to changes in the stress of
$\pm 0.026\ {\rm dyne/cm}$.
Therefore, when measuring the changes in the stress, changes of
$\pm 0.026\ {\rm dyne/cm}$ were ignored, providing
a lower limit on the size of the stress drops.

The bubble raft is generated by flowing nitrogen through a
solution of 44\% by weight glycerine, 28\% by weight water
and 28\% by weight Miracle Bubbles (Imperial Toy Corp.).
The bubble size is
fixed by the pressure and needle diameter. For the experiments
reported on here, three different size bubbles
(2 mm, 3 mm, 5 mm) were used, with approximately
500 bubbles of each size. The experiments were carried out over
two hours. The coarsening of the foam without shear was monitored
during this time, and no significant coarsening or stress drops
due to coarsening were observed.
However, after two-hours, the stability of the foam deteriorated
rapidly, and shear resulted in severe rupture of bubbles.

Figure 1 is an image of a section of the bubble raft between the
two cylinders. The bubble raft was monitored for slippage at both
the outer and inner cylinder. For all of the experiments reported
here, there was no slip between the first row of bubbles and the
corresponding cylinder. Also, it was clear from the images that
the stress drop events corresponded to rearrangements of the
bubbles in the bulk of the foam. The average azimuthal velocity
distribution was consistent with what one would expect for
Couette flow. This is illustrated in Fig. 2 for an angular
rotation of $0.01\ {\rm s^{-1}}$.  The dashed
line is the theoretical velocity profile for a Newtonian fluid,
and the solid curve is the theoretical velocity profile for
a shear-thinning fluid with a viscosity given by
$\eta = m\dot{\gamma}^{n-1}$, where $m$ is a constant,
$\dot{\gamma}$ is the shear rate, and $n = 1/3$ \cite{BAH77}.
It should be noted
that the bubble model predicts shear-thinning with an exponent
of approximately $1/3$ \cite{LL00}. The fact that the velocity
profile is consistent with the expected shear-thinning profile, and
not the velocity profile of the water substrate, is strong evidence
that the foam is acting independent of the water. Because the rate
of strain varies as a function of radius, all reported values of
for rates of strain are taken at the inner cylinder. Finally, it should
be noted that this is very different behavior from that observed
for two-dimensional foams confined between glass plates. In this
case, exponential decay of the velocity profile was
observed \cite{DTM01}.

Figure 3 shows a typical response of the bubble raft to shear. The
stress on the inner cylinder is shown as a function of the applied
strain for a rate of strain of $3.1 \times 10^{-3}\ {\rm s^{-1}}$.
The two key features to notice are the
initial elastic region where the stress increases linearly with
strain and the subsequent region in which flow occurs. The
intermittent nature of the flow is obvious in this plot. The yield
strain was always of order 1 for all rates of strain.

Figure 4 shows the distribution of the stress drops for three
different rates of strain. Here the size of the drop is normalized
by the average stress per bubble. Also, as discussed above, any
change in stress of $\pm 0.026\ {\rm dyne/cm}$ was ignored and does not enter into
the plotted distribution. The distribution of stress drops is not consistent with a
power-law for all length scales; however, for small stress drops,
the distribution is consistent with a power law with an exponent
of -0.8. For comparison, the exponent predicted for the bubble
model is $-0.70 \pm 0.05$ \cite{D97}.
Also, as predicted by the bubble model \cite{TSDKLL99},
the distribution is essentially independent of
the rate of strain for small stress drops, and the large stress
drop cutoff is weakly dependent on the rate of strain. This is
difficult to see from Fig. 4, but is made clear by considering the
average stress drop size. For example, for a rate of strain of
$0.031\ {\rm s^{-1}}$, the average stress drop (normalized
by the average stress per bubble) is 27.9; whereas, it is
47.0 for a rate of strain of $0.31\ {\rm s^{-1}}$.

We have conclusively demonstrated that the general features of the
distribution of stress drops observed in the bubble raft are
in agreement with the simulations reported in Ref. \cite{D97} and
\cite{TSDKLL99}. Specifically, the cutoff in the distribution for
large stress drops is clear. Also, the viscosity of the bubble
raft is consistent with a shear-thinning fluid with an
exponent of 1/3. Finally, the distribution of stress drops,
particularly for the small stress-drops, is essentially independent
of the rate of strain for the values studied here. This strongly suggest
the existence of a quasi-static limit. Such a limit is particularly
important for our geometry where the rate of strain varies across
the system.

A number of important questions
remain. First, what is the correspondence between the size
of the stress drops and the number of bubbles that exchange
neighbors? It is still possible that the larger stress drops are
due to rearrangements involving a wide distribution of the number
of bubbles, as reported in \cite{JSSAG99}. Therefore,
``system-wide'' events are still a
possibility. The current imaging system was not appropriate for
careful tracking of bubbles throughout the system.
Future improvements in the imaging of
this system will enable this question to be addressed. Second, a
detailed study of the flow behavior of bubble rafts as a function
of both disorder (polydispertiy) and area fraction will provide
further tests of the various two-dimensional models of foams. Of
particular importance is the behavior as a function of area fraction
to distinguish further between the predictions of the quasi-static
simulations \cite{WBHA92,HWB95} and the bubble model \cite{D97,TSDKLL99}.

The authors acknowledge funding from NSF grant CTS-0085751. M. Dennin
acknowledges additional funding from the Research Corporation
and Alfred P. Sloan Foundation.
J. Lauridsen was supported by NSF Research Experience for
Undergraduates grant PHY-9988066 and a grant from UCI-UROP
program. The authors thank Mingming Wu,
Andrea Liu, and Doug Durian for useful conversations.

\begin{figure}[htb]
\epsfxsize = 3.0in
\centerline{\epsffile{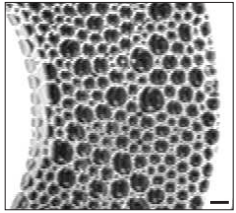}}
\caption{This is an image of one section of a typical bubble raft.
Part of both the inner and outer cylinder are visible. The black scale-bar
in the lower right corner is 3.6 mm.}
\end{figure}

\begin{figure}[htb]
\epsfxsize = 3.0in
\centerline{\epsffile{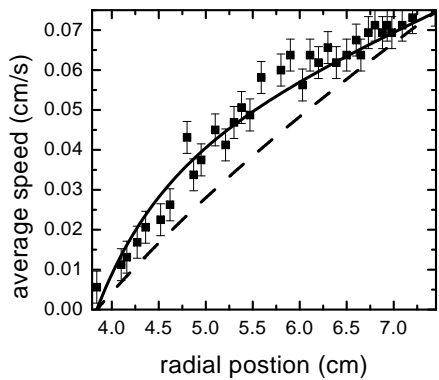}}
\caption{Plot of average azimuthal speed of the bubbles versus radial position
(solid symbols). The solid line is the theoretical velocity profile for a
shear-thinning fluid with n = 1/3 (see the text for details), and the dashed
line is the theoretical profile for a Newtonian fluid.}
\end{figure}

\begin{figure}[htb]
\epsfxsize = 3.0in
\centerline{\epsffile{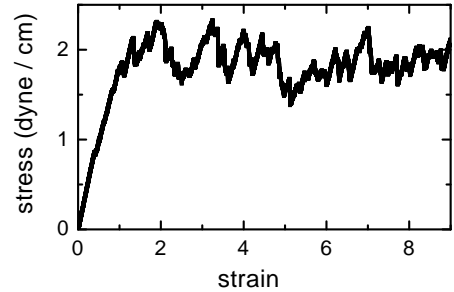}}
\caption{Plot of the stress versus strain for
 a rate of strain of $3.1 \times 10^{-3}\ {\rm s^{-1}}$.}
\end{figure}

\begin{figure}[htb]
\epsfxsize = 3.0in
\centerline{\epsffile{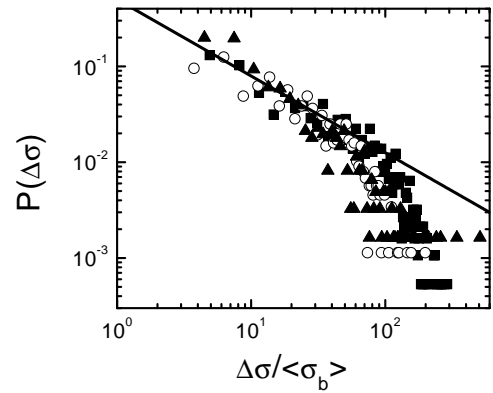}}
\caption{Distribution of stress drops for three different rates of strain.
Solid triangles are a rate of strain of $0.031\ {\rm s^{-1}}$,
solid squares are a rate of strain of $0.31\ {\rm s^{-1}}$, and 
open circles are for a rate of strain of $0.48\ {\rm s^{-1}}$. The solid
line has a slope of -0.8 and is a guide to the eye.}
\end{figure}


\begin{references}

\bibitem{REV} There is a vast literature on foams and their
flow behavior. Some reviews are: D. Weaire and S. Hutzler,
{\it The Physics of Foams}, (Claredon Press, Oxford, 1999), J. Stavans, Rep. Prog.
Phys. {\bf 56}, 733 (1993), and A. M. Kraynik, Ann. Rev. Fluid
Mech. {\bf 20}, 325 (1988).

\bibitem{CWBC98} M. E. Cates, J. P. Wittmer, J.-P. Bouchaud, and P. Claudin,
Phys. Rev. Lett. {\bf 81}, 1841 (1998).

\bibitem{LN98} A. J. Liu and S. R. Nagel, Nature {\bf 396}, 21 (1998).

\bibitem{TPCSW01} V. Trappe, V. Prasad, L. Cipelletti, P. N. Segre,
D. A. Weitz, Nature {\bf 411}, 772 (2001).

\bibitem{ITPJamming} {\it Jamming and Rheology}, eds. A. J. Liu and S. R. Nagel,
(Taylor and Francis Group, 2001).

\bibitem{KNN89} K. Kawasaki, T. Nagai, and K. Nakashima, Phil. Mag. B
{\bf 60}, 399 (1989).

\bibitem{KOKN95} K. Kawasaki, T. Okuzono, T. Kawakatsu, and T. Nagai,
Proc. Int. Workshop of Physics of Pattern Formation ed. S. Kai
(Singapore: World Scientific, 1992).

\bibitem{OK95} T. Okuzono and K. Kawasaki, Phys. Rev. E {\bf 51}, 1246 (1995).

\bibitem{D95} D. J. Durian, Phys. Rev. Lett. {\bf 75}, 4780 (1995).

\bibitem{D97} D. J. Durian, Phys. Rev. E {\bf 55}, 1739 (1997).

\bibitem{WBHA92} D. Weaire, F. Bolton, T. Herdtle, and H. Aref,
Phil. Mag. Lett. {\bf 66}, 293 (1992).

\bibitem{HWB95} S. Hutzler, D. Weaire, and F. Bolton, Phil. Mag. B {\bf 71}, 277 (1995).

\bibitem{JSSAG99} Y. Jiang, P. J. Swart, A. Saxena, M. Asipauskas, and J. A. Glazier,
Physical Review E {\bf 59}, 5819 (1999).

\bibitem{TSDKLL99} S. Tewari, D. Schiemann, D. J. Durian, C. M. Knobler, S. A. Langer,
and A. J. Liu, Phys. Rev. E {\bf 60}, 4385 (1999).

\bibitem{DK97} M. Dennin and C. M. Knobler, Phys. Rev. Lett. {\bf
78}, 2485 (1997).

\bibitem{KE99} A. A. Kader and J. C. Earnshaw, Phys. Rev. Lett.
{\bf 82}, 2610 (1999).

\bibitem{GD95} A. D. Gopal and D. J. Durian, Phys. Rev. Lett. {\bf
75}, 2610 (1995).

\bibitem{AK79} A. S. Argon and H. Y. Kuo, Mat. Sci. and Eng. {\bf 39},
101 (1979).

\bibitem{MGC89} D. Mazuyer, J. M. Georges, and B. Cambou, J. Phys. France
{\bf 49}, 1057 (1989).

\bibitem{app} R. S. Ghaskadvi and M. Dennin, Rev. Sci. Instr. {\bf 69},
3568 (1998).

\bibitem{BAH77} R. B. Bird, R. C. Armstrong, and O. Hassuage,
{\it Dynamics of Polymer Liquids} (Wiley, Newyork, 1977).

\bibitem{LL00} S. A. Langer and A. J. Liu, Europhys. Lett. {\bf 49}, 68
(2000).

\bibitem{DTM01} G. Debr\'{e}geas, H. Tabuteau, and J. -M. di Meglio,
Phys. Rev. Lett. {\bf 87}, 178305 (2001).

\end{references}
\end{document}